\documentclass[reprint,letterpaper,superscriptaddress, pra]{revtex4-1}
\usepackage{amssymb}

\usepackage{graphicx}
\usepackage{amsfonts}
\usepackage{amsmath}
\usepackage{grffile}
\usepackage{MnSymbol,wasysym}

\renewcommand{\vec}[1]{\mathbf{#1}}

\newcommand{\tm}[1]{\textrm{#1}}

\begin{document}

\title{Bogoliubov theory of a Bose-Einstein condensate of rigid rotor
molecules}
\author{Joseph C.~Smith}
\affiliation{Department of Physics and Astronomy, Western Washington
University, Bellingham, Washington 98225, USA}
\author{Seth T.~Rittenhouse}
\affiliation{Department of Physics, The United States Naval Academy,
Annapolis, Maryland 21402, USA}
\author{Ryan M.~Wilson}
\affiliation{Department of Physics, The United States Naval Academy,
Annapolis, Maryland 21402, USA}
\author{Brandon M.~Peden}
\email[Contact author: ]{brandon.peden@wwu.edu}
\affiliation{Department of Physics and Astronomy, Western Washington
University, Bellingham, Washington 98225, USA}

\date{\today}

\begin{abstract}
We consider a BEC of rigid rotor molecules confined to quasi-2d through
harmonic trapping. The molecules are subjected to an external electric field
which polarizes the gas, and the molecules interact via dipole-dipole interactions.
We present a description of the ground state and low-energy excitations of
the system including an analysis of the mean-field energy, polarization, and stability.
Under large electric fields the gas becomes fully polarized and we reproduce
a well known density-wave instability which arises in polar BECs. Under smaller applied electric fields the gas develops an in-plane
polarization leading to the emergence of a new global instability as the
molecules ``tilt''.  The character of these instabilities is clarified by means of
momentum-space density-density structure factors. A peak at zero momentum in the spin-spin structure factor
for the in-plane component of the polarization indicates that the tilt instability is a global phonon-like instability.
\end{abstract}

\pacs{Valid PACS appear here}
\maketitle

\section{\label{sec:intro}Introduction}

The first experimental realization of a high phase-space density,
ultracold gas of polar molecules was achieved in 2010 in a gas of KrB molecules \cite{Ospelkaus10,Yan13}.
Since then, researchers have achieved a high phase space density gas at ultra-low temperatures in a number of other bialkali molecular systems, including NaK \cite{Park2015,Seesselberg18}, RbCs \cite{Takekoshi14,Molony14}, NaRb \cite{guo16}, and LiNa \cite{rvachov17}. With long-range, anisotropic dipole-dipole interactions, cold-molecular systems are ideal for realizing a wide range of interesting systems, with researchers proposing spin lattice models \cite{Moses2017,Micheli06,Osterloh07,Buchler07,Gorshkov11} and topological phases \cite{Cooper09,Yao13} as well as realizing quantum chemistry in the ultracold regime \cite{Carr09,Bohn17}. The recent advancement in which a gas of KRb
molecules (which are composite fermions) was cooled far enough with a high
enough number density to achieve degeneracy has further spurred on this
excitement \cite{DeMarco19}.   Further advancements in the laser cooling of $X^2\Sigma$ molecules \cite{Barry14,Norrgard16,Steinecker16,Truppe17} and even more complex molecular structures (such as CaOH and CaOCH$_3$ \cite{Kozyryev19}) has been reported.  With this rapidly expanding experimental progress we can assume that the achievement of a Bose-Einstein condensate (BEC) of polar molecules is on the horizon, particularly in light of recent proposals to mitigate losses in such systems \cite{Lassabliere2018,Karman18}.

In this work we are interested in the effect of internal degrees of freedom
(encoded in the rotational levels of the molecules) on the behavior of a BEC
of polar molecules confined to a quasi-2D geometry with strong dipole-dipole interactions (ddi). In previous work, researchers predicted that when these degrees of freedom are removed using a strong
external field to polarize the molecules in the lab frame, the
quasi-particle dispersion of the quasi-2D dipolar gas develops a roton-maxon
structure \cite{Santos03}. With large enough interaction strength, the gas becomes
dynamically unstable against a density wave instability \cite{Ronen06,Wilson2008a,Ticknor11,Bisset13}. The signatures of this density wave roton mode have recently been observed in a quasi-1D gas of erbium atoms interacting via strong magnetic ddi \cite{Chomaz18}.  Later, researchers predicted that when the internal structure of molecules is
incorporated at the level of a two-state approximation, a different
instability, dominated by polarization fluctuations, emerges at low external
field and high interaction strengths \cite{Wilson14,Peden2015}. Here we extend this
work to include multiple internal molecular configurations in the form of
quantized rotational states.
Using a rigid
rotor approximation, in which we consider the full manifold of rotational angular
momentum states, we build in a robust model for quantum polarizability.
Interactions between individual molecules with the trapping potential,
external electric field, and other molecules create many competing energy
scales in this system. This leads to rich and diverse behavior not only in
the ground state, but also the low energy excitations. 

To emphasize the role
of the long-range, state-dependent, anisotropic dipole-dipole interaction we
set all short range interactions to zero. Recently, it has been noted that
the complex nature of molecule-molecule interactions at short range can play
a significant role in the loss dynamics of the dipolar system.  Molecules such as KRb can be unstable under collisions to processes like $%
\text{KRb}+\text{KRb}\rightarrow \text{Rb}_{2}+\text{K}_{2}+\text{KE}$ causing reactive two-body losses from
the cold gas \cite{Ospelkaus10,Yan13}. In addition, even
nonreactive species of molecules can have a two-body loss rate similar to
reactive species \cite{guo16}. It has been suggested that this is due to the complex
meta-stable bound state structure with a large density of states present in
many molecule-molecule collisions which can cause them to have a long
average scattering time, the so-called ``sticky-molecule'' effect \cite{Mayle13}.
Recently, it has been observed that strong two-body losses might be driven by optical excitations of the short-range few-body complex to anti-trapping electronic states \cite{gregory20}.
Such short range losses might be shielded with an induced repulsive isotropic interaction \cite{Karman18}. The incorporation of loss
mechanisms such as these are the subject of ongoing study and are neglected
in this work.

This paper is divided into $4$ sections. The underlying theory is presented
in Sec.~\ref{sec:theo}, in which we present both the single and many-body
Hamiltonian. We derive the fluctuation Hamiltonian that governs the behavior of the low-energy excitations by way of Bogoliubov-de Gennes theory. We conclude this section with methods for obtaining
static structure factors from the two-point correlation functions. In
Sec.~\ref{sec:MFGS}, we employ mean-field theory and numerically minimize
the ground state energy with respect to the internal state amplitudes. We
characterize the response of the gas to changes in the interaction and
electric field strength through the polarization, polarizability and the
energy.

The analysis of the low energy excitations is broken up into two sections.
In Sec.~\ref{sec:LEE}, we diagonalize the fluctuation Hamiltonian and
obtain the quasiparticle dispersion relations. Using these, we
characterize the quasiparticle spectrum. In our analysis, we
identify three distinct ways in which the gas becomes susceptible to
mesoscopic fluctuations in the polarization and the density. As the gas
gives rise to these fluctuations the gas also becomes susceptible to
instabilities. In the last section of this paper, Sec.~\ref{sec:physicalcharacterization}, we
compute the momentum space density-density and spin-spin structure factors
and use them to characterize the nature of the instabilities seen in
Sec.~\ref{sec:LEE}. We identify the emergence of
previously predicted density and spin wave instabilities and the emergence
of a new long-wavelength phonon instability associated with the spontaneous symmetry breaking of the azimuthal symmetry which occurs as the molecular polarization tilts relative to the external field.

\section{\label{sec:theo}Theory}

We consider a gas of diatomic, hetero-nuclear, rigid-rotor
molecules confined to two dimensions via harmonic trapping. The molecules
experience an external electric field $\vec{E}$, directed perpendicular to
the plane of trapping, which polarizes the molecules, causing them to
interact via dipole-dipole interactions. Here, we are modeling the effects
that the internal structure of the molecules has on the many-body behavior
of the gas. We build in a microscopic, quantum mechanical
treatment of the internal structure of the molecule by including a truncated
set of lab-frame rotational states. We thereby fully treat the effects on the polarization of the gas of the external electric field and of the mean-field dipole field acting back on a single molecule on the gas. In addition, this model allows a complete description of the spin-exchange between the internal (molecular) and external (center-of-mass) degrees of freedom, which is the key physics involved in a novel tilt instability that we predict to occur in the low-field limit (see Sec.~\ref{sec:physicalcharacterization}).

By way of Bogoliubov-de Gennes theory, we derive the ground state energy and
fluctuation Hamiltonian, which we diagonalize in order to find both the
dispersion relations characterizing the low-lying excitations of the gas and
spin and density structure factors characterizing the mesoscopic behavior of
the BEC.

\subsection{Single-molecule theory}

The Hamiltonian of a single rigid-rotor molecule in the presence of an
external electric field $\vec{E}$ is given by 
\begin{align}
\hat{H}_{\mathrm{mol}}=\frac{B}{\hbar^{2}}\hat{J}^{2}-\hat{\vec{d}}\cdot 
\vec{E},
\end{align}
where $B$ is the rotational constant, $\vec{E}$ is the external electric
field, $\hat{J}$ is the total rotational angular momentum, and $\hat{\vec{d}}
$ is the dipole moment operator for the molecule. The first term represents
the kinetic energy of rotation, and the second term represents the
interaction of the dipole moment of the molecule with the external field.
We note that we have assumed that all vibrational and electronic excitations are energetically inaccessible and hence frozen out.
We work in the basis of lab-frame angular momentum states $\left\vert
jm\right\rangle $, in which $\hat{J}^2$ is diagonal, given by 
\begin{align}
\langle jm|\hat{J}^{2}|j^{\prime}m^{\prime}\rangle
=\delta_{jj^{\prime}}\delta_{mm^{\prime}}\hbar^{2}j\left( j+1\right) .
\end{align}
The spherical-tensor components of the dipole moment operator $\hat{\vec{d}}$
can be written in terms of $3j$ symbols as 
\begin{align}
\langle j,m|\hat{d}_{\mu}|j^{\prime}m^{\prime}\rangle&=d\left( -1\right) ^{m}%
\sqrt{\left( 2j+1\right) \left( 2j^{\prime}+1\right) }  \notag \\
&\quad\mbox{}\times \left( 
\begin{array}{ccc}
j & 1 & j^{\prime} \\ 
-m & \mu & m^{\prime}%
\end{array}
\right) \left( 
\begin{array}{ccc}
j & 1 & j^{\prime} \\ 
0 & 0 & 0%
\end{array}
\right) ,
\end{align}
where $d$ is a matrix element in vibrational states and represents the
magnitude of the body-fixed molecular dipole moment \cite{Bohn2013}.

\subsection{Many-body Hamiltonian}

The full many-body Hamiltonian is given by%
\begin{align}
\hat{\mathcal{H}} & =\int d^{3}r\hat{\Psi}^{\dagger}\left( \vec{r}\right) \left( \hat{H}_{%
\mathrm{CM}}\left( \vec{r}\right) +\hat{H}_{\mathrm{mol}}\right) \hat{\Psi}%
\left( \vec{r}\right)  \notag \\
& \quad{}\mbox{}+\frac{1}{2}\int d^{3}r_{1}\hat{\Psi}^{\dagger}\left( \vec{r}%
_{1}\right) \hat{\Psi}^{\dagger}\left( \vec{r}_{2}\right) \hat {U}\left( 
\vec{r}_{1}-\vec{r}_{2}\right) \hat{\Psi}\left( \vec{r}_{2}\right) \hat{\Psi}%
\left( \vec{r}_{1}\right) .
\end{align}
Here, $\hat{H}_{\mathrm{CM}}\left( \vec{r}\right) $ is the single-particle
Hamiltonian for the center of mass motion, given by%
\begin{align}
\hat{H}_{\mathrm{CM}}\left( \vec{r}\right) =\left(-\frac{\hbar^{2}}{2m}\nabla^{2}+\frac{1}{2}m\omega^{2}z^{2}\right)\hat{I},
\end{align}
where the second term represents the harmonic trapping,
and $\hat{I}$ is the identity operator on the internal molecular states.
$\hat{U}\left( 
\vec{r}_{1}-\vec{r}_{2}\right) $ is the dipole-dipole interaction, given by%
\begin{align}
\hat{U}\left( \vec{r}\right) =\frac{1}{r^{3}}\left( \hat{\vec{d}}_{1}\cdot\hat{%
\vec{d}}_{2}-\left( \hat{\vec{d}}_{1}\cdot\hat{\vec{r}}\right) \left( \hat{%
\vec{d}}_{2}\cdot\hat{\vec{r}}\right) \right) .
\end{align}
We expand the field operators in a single-molecule basis $\left\{ \left\vert
n\right\rangle \right\} $---in practice, this is either $\left\{ \left\vert
jm\right\rangle \right\} $ or the eigenbasis of the single-molecule
Hamiltonian---as%
\begin{align}
\hat{\Psi}\left( \vec{r}\right) =\sum_{n}\hat{\psi}_{n}\left( \vec {r}%
\right) \left\vert n\right\rangle ,
\end{align}
yielding
\begin{subequations}
\label{eqn:FullManyBodyHamiltonian}
\begin{align}
\hat{\mathcal{H}} & =\hat{\mathcal{H}}_{0}+\hat{\mathcal{H}}_{1}+\hat{\mathcal{V}}-\mu\hat{\mathcal{N}},
\end{align}
where $\hat{\mathcal{H}}_{0}$, given by 
\begin{align}
\hat{\mathcal{H}}_{0}=\sum_{n}\int d^{3}r\hat{\psi}_{n}^{\ast}\left( \vec{r}\right) 
\hat{H}_{\mathrm{CM}}\left( \vec{r}\right) \hat{\psi}_{n}\left( \vec {r}%
\right) ,
\end{align}
is the center-of-mass Hamiltonian, $\hat{\mathcal{H}}_{1}$, given by%
\begin{align}
\hat{\mathcal{H}}_{1}=\sum_{mn}\left\langle m\right\vert \hat{H}_{\mathrm{mol}%
}\left\vert n\right\rangle \int d^{3}r\hat{\psi}_{m}^{\ast}\left( \vec {r}%
\right) \hat{\psi}_{n}\left( \vec{r}\right) ,
\end{align}
is the Hamiltonian for the internal molecular states, $\hat{\mathcal{V}}$, given by%
\begin{align}
\hat{\mathcal{V}}&=\frac{1}{2}\sum_{m_{1}m_{2}n_{1}n_{2}}\int d^{3}r_{1}\hat{\psi}%
_{m_{1}}^{\dagger}\left( \vec{r}_{1}\right) \hat{\psi}_{m_{2}}^{\dagger
}\left( \vec{r}_{2}\right) \hat{\psi}_{n_{1}}\left( \vec{r}_{1}\right) \hat{%
\psi}_{n_{2}}\left( \vec{r}_{2}\right)  \notag \\
&\quad\mbox{}\times \left\langle m_{1}\right\vert \left\langle
m_{2}\right\vert \hat{U}\left( \vec{r}_{1}-\vec{r}_{2}\right) \left\vert
n_{2}\right\rangle \left\vert n_{1}\right\rangle ,
\end{align}
\end{subequations}
is the interaction Hamiltonian, $\hat{\mathcal{N}}$ is the number operator, and we have included a chemical potential $\mu$ in order to work in the grand-canonical ensemble, in preparation for
the Bogoliubov de-Gennes analysis to follow.

\subsection{Bogoliubov theory}

We make the assumption that the ground-state wave function factorizes into
axial ($z$) and transverse (${\boldsymbol{\rho}}$) components, which is a
good approximation provided that the trapping is sufficiently tight. (For
further justification of this approximation, see Ref.~\cite{Peden2015}.) In
addition, since we treat the system as being free in-plane, we can make the
approximation that the ground state has a uniform density. We then expand
the state-indexed field operators $\hat{\psi}_{n}\left( \vec {r}\right) $ as
a sum of condensate and fluctuation terms, given by 
\begin{align}
\hat{\psi}_{n}\left( \vec{r}\right) =\frac{f_{n}\left( z\right) }{\sqrt {A}}%
\left( \sqrt{N}\alpha_{n}+\sum_{\vec{k}\neq0}e^{i\vec{k}\cdot{\boldsymbol{%
\rho}}}\hat{a}_{\vec{k},n}\right).
\label{eqn:Expansion}
\end{align}
Here, $A$ is the in-plane area occupied by the gas, $N$ is the total number
of molecules, $f_{n}\left( z\right) $ is the axial wave function, $%
|\alpha\rangle$, given by 
\begin{align}
\left\vert \alpha\right\rangle =\sum_{n}\alpha_{n}\left\vert n\right\rangle,
\end{align}
is the internal-state wave function, and $\hat{a}_{\vec{k},n}$ is the
annihilation operator for a particle in state $\left\vert n\right\rangle $
with in-plane momentum $\vec{k}$.

We assume that the
center-of-mass motion is the same for particles in different states and
employ a Gaussian ansatz, 
\begin{align}
f_{n}\left( z\right) =\frac{1}{\sqrt{\pi\sqrt{l}}}e^{-z^{2}/2l^{2}}, 
\label{eqn:GaussianAnsatz}
\end{align}
where $l=\sqrt{\hbar/m\omega}$ is the oscillator length for the harmonic
trapping.
These approximations have been shown to be only qualitatively accurate~\cite{Peden2015}.

Since the gas is in a BEC state, the $\vec{k}=0$ state is macroscopically occupied, in which case the terms that are quartic in the $\vec{k}\neq0$ raising and lowering operators are small compared to the quadratic terms. In addition, the odd-order terms vanish in the mean-field ground state. We therefore keep only those terms in the expansion that are constant or quadratic in
the creation and annihilation operators. We collect the former into the
ground-state energy functional $K_0$, given by 
\begin{align}
\frac{K_{0}}{N} & =-\mu\left\langle \alpha|\alpha\right\rangle +\frac {%
\hbar\omega}{2}\left\langle \alpha|\alpha\right\rangle  \notag \\
&\quad\mbox{}+\left\langle \alpha\right\vert \hat{H}_{\mathrm{mol}%
}\left\vert \alpha\right\rangle +\frac{1}{2}\left\langle
\alpha\alpha\right\vert \hat{\Lambda}_{0}\left\vert
\alpha\alpha\right\rangle ,  \label{eqn:GroundStateEnergyFunctional}
\end{align}
and the latter into the fluctuation Hamiltonian $\hat{K}_2$. The operator $%
\hat{\Lambda}_0$ is the zero-momentum limit of the interaction operator
integrated over the spatial degrees of freedom, given by 
\begin{align}
\hat{\Lambda}_{\vec{k}}&=\frac{N}{A^{2}}\int d^{3}r_{1}\left\vert f\left(
z_{1}\right) \right\vert ^{2}e^{-i\vec{k}\cdot{\boldsymbol{\rho}}_{1}} 
\notag \\
&\quad\mbox{}\times \int d^{3}r_{2}\hat{U}\left( \vec{r}_{1}-\vec{r}%
_{2}\right) \left\vert f\left( z_{2}\right) \right\vert ^{2}e^{i\vec{k}\cdot{%
\boldsymbol{\rho}}_{2}}.  \label{eqn:IntegratedInteractionOperator}
\end{align}
Details of this derivation and of the Bogoliubov diagonalization procedure
are outlined in Appendix \ref{app:Bogoliubov}. Here, we note that we
minimize $K_0$ with respect to $\alpha_n$ to find the
mean field ground state, and we then diagonalize $\hat{K}_2$ via a canonical
transformation of the annihilation operators, given by~\cite{MingWen2009}
\begin{align}
\hat{a}_{\vec{k},n} = \sum_{m}\left(U_{\vec{k},nm}\hat{b}_{\vec{k},n}+V_{-%
\vec{k},nm}\hat{b}_{-\vec{k},m}^{\dagger}\right),
\end{align}
where the $U$ and $V$ matrices are defined in the appendix. This results in
a fluctuation Hamiltonian of the form
\begin{align}
\hat{K}_{2}=\sum_{\vec{k},n}\frac{\Omega_{\vec{k},n}}{2}\left( \hat{b}_{\vec{%
k},n}^{\dagger}\hat{b}_{\vec{k},n}+\hat{b}_{-\vec{k},n}\hat{b}_{-\vec{k}%
,n}^{\dagger}\right) ,
\end{align}
where $\Omega_{\vec{k},n}$
are the quasi-particle dispersion relations.

\subsection{Dipole-dipole interactions}

The diagonalization procedure above can be performed numerically as long as
the integrated interaction operator $\hat{\Lambda}_{\vec{k}}$ can be
computed. It turns out that this can be done analytically, and we present
the full details of the derivation in Appendix \ref%
{app:MultipoleInteractions}. Here, we quote the result. The integrated
dipole-dipole interaction term can be written as 
\begin{align}
\hat{\Lambda}_{\vec{k}} & =\hbar\omega g_{\text{d}}\left( 2\hat{d}_{0}\otimes\hat{d}_{0}+%
\hat{d}_{1}\otimes\hat{d}_{-1}+\hat{d}_{-1}\otimes\hat{d}_{1}\right) F\left( 
\frac{kl}{\sqrt{2}}\right)  \notag \\
& \quad\mbox{}+\hbar\omega g_{\text{d}}\left( e^{i2\phi}\hat{d}_{-1}\otimes\hat{d}%
_{-1}+e^{-i2\phi}\hat{d}_{1}\otimes\hat{d}_{1}\right) \left( 1-F\left( \frac{%
kl}{\sqrt{2}}\right) \right), \label{eqn:integratedddionoperator}
\end{align}
where $g_{\text{d}}$, given by%
\begin{align}
g_{\text{d}}=\frac{Nd^{2}}{Al\hbar\omega}\frac{\sqrt{8\pi}}{3},
\end{align}
is an effective interaction strength, $\hat{d}_{\mu}$ are the spherical
components of the dipole moment operator, and 
\begin{align}
F\left( x\right) & =1-\frac{3}{2}\sqrt{\pi}xe^{x^{2}}\mathrm{erfc}\left(
x\right) , \\
e^{i\phi} & =\frac{k_{x}+ik_{y}}{k}.
\end{align}

Because the components $\hat{d}_{\mu}$ of the dipole moment operator do not commute with each other, the mean-field ground state may display non-azimuthal symmetry despite the azimuthal symmetry of the system about the direction of the electric field (which is aligned with the trap axis). This physics manifests as an instability in which the net dipole moment of the gas ``tilts''---that is, develops an in-plane component---which breaks the azimuthal symmetry of the system.

\subsection{Static structure factors}

The behavior of the quasi-particle fluctuations can be characterized by way
of static structure factors. We define the structure factors by way of the
two-point correlation functions, given by~\cite{Symes2014}
\begin{align}
G_{\hat{w}_{1}\hat{w}_{2}}^{\left( 2\right) }\left( {\boldsymbol{\rho}}%
\right) =\left\langle N\left[ \delta\hat{w}_{1}\left( {\boldsymbol{\rho}}%
\right) \delta \hat{w}_{2}\left( 0\right) \right] \right\rangle ,
\end{align}
where $\hat{w}_{j}$ is a single-particle operator, and $N\left[ \cdots\right]
$ denotes normal ordering. The structure factors and correlation functions
are related via%
\begin{align*}
S_{\hat{w}_{1}\hat{w}_{2}}\left( \vec{k}\right) =\frac{1}{n}\int d^{2}\rho
e^{-i\vec{k}\cdot{\boldsymbol{\rho}}}G_{w_{1}w_{2}}^{\left( 2\right) }\left( 
{\boldsymbol{\rho}}\right) +\left\langle \alpha\right\vert
W_{1}W_{2}\left\vert \alpha\right\rangle ,
\end{align*}
where $W_{j}$ is the matrix forms of $\hat{w}_{j}$, and it can be shown that 
\begin{align}
S_{\hat{w}_{1}\hat{w}_{2}}\left( \vec{k}\right) & =\sum_{j}\left(
\left\langle \alpha\right\vert W_{1}U_{\vec{k}}\left\vert j\right\rangle
+\left\langle j\right\vert V_{\vec{k}}^{\dagger}W_{1}\left\vert \alpha
\right\rangle \right)  \notag \\
& \quad\mbox{}\times\left( \left\langle \alpha\right\vert W_{2}V_{\vec{k}%
}\left\vert j\right\rangle +\left\langle j\right\vert U_{\vec{k}}^{\dagger
}W_{2}\left\vert \alpha\right\rangle \right) .
\end{align}
Finally, we define normalized structure factors $s_{\hat{w}_{1}\hat{w}_{2}}(%
\vec{k})$ that are normalized to the long-distance (uncorrelated) values,
e.g. 
\begin{align}
s_{\hat{w}_{1}\hat{w}_{2}}(\vec{k}) &= \frac{S_{\hat{w}_{1}\hat{w}_{2}}(\vec{%
k})}{\left\langle \alpha\right\vert W_{1}W_{2}\left\vert \alpha\right\rangle}%
.  \label{eqn:StructureFactor}
\end{align}
In particular, we will be interested in the density structure factor $\hat{s}%
_n$, where $\hat{w}=n\hat{I}$ and $\left\langle \alpha\right\vert
W_{1}W_{2}\left\vert \alpha\right\rangle =1$, the $z$-component polarization
structure factor $\hat{s}_{z}$, where $\hat{w} = \hat{d}_z$, and the
in-plane polarization structure factor $s_{\perp}$, defined by 
\begin{align}
s_{\perp} &= \frac{S_{\hat{d}_x\hat{d}_x}+S_{\hat{d}_y\hat{d}_y}}{%
\left\langle \alpha\right\vert \hat{d}_x^2\left\vert\alpha\right\rangle
+\left\langle \alpha\right\vert \hat{d}_y^2\left\vert\alpha\right\rangle}.
\label{eqn:StructureFactorPerp}
\end{align}

\section{Mean-Field Ground State}

\label{sec:MFGS}

In this section, we discuss the behavior of the mean-field ground state. We
numerically solve for the mean-field ground state $\left|{\alpha}%
\right\rangle$ by minimizing the energy functional (Eq.~\ref%
{eqn:GroundStateEnergyFunctional}). We use $\left|{\alpha}\right\rangle$ to
compute the polarization $\vec{p}$, polarizability tensor $\boldsymbol{\alpha}$, and energy $E$
of the ground state as a function of the effective density-dependent
interaction strength $g_{\text{d}}$ and effective electric field strength $%
\beta=dE/B$. In the following, we assume
that the rotational constant $B$ is equal to the trap energy $\hbar\omega$. This choice is well beyond experimental accessibility, but we also obtain the qualitative behavior discussed for more reasonable choices of $B/\hbar\omega$. We discuss the effects of changing this ratio and the experimental implications in Sec.~\ref{sec:Conclusion}.

We identify a second-order phase transition between a state in which the
dipole moment is aligned with the external field and a state in which the
polarization has a non-zero component in the trapping plane. The physics underlying these phases is as follows.
The local field sampled by each molecule is the combination of the external field and the mean-field dipole field created by all other molecules in the gas.  Since the dipole field of the other molecules is locally anti-aligned with the
external field at the position of each molecule, these two contributions compete with each other, and they therefore self-consistently determine the net dipole moment (polarization) of the gas. This competition can be represented in a phase diagram of sorts, shown below as the polarization as a function of field strength and interaction strength.

\begin{figure}[t!]
\includegraphics[width=86mm]{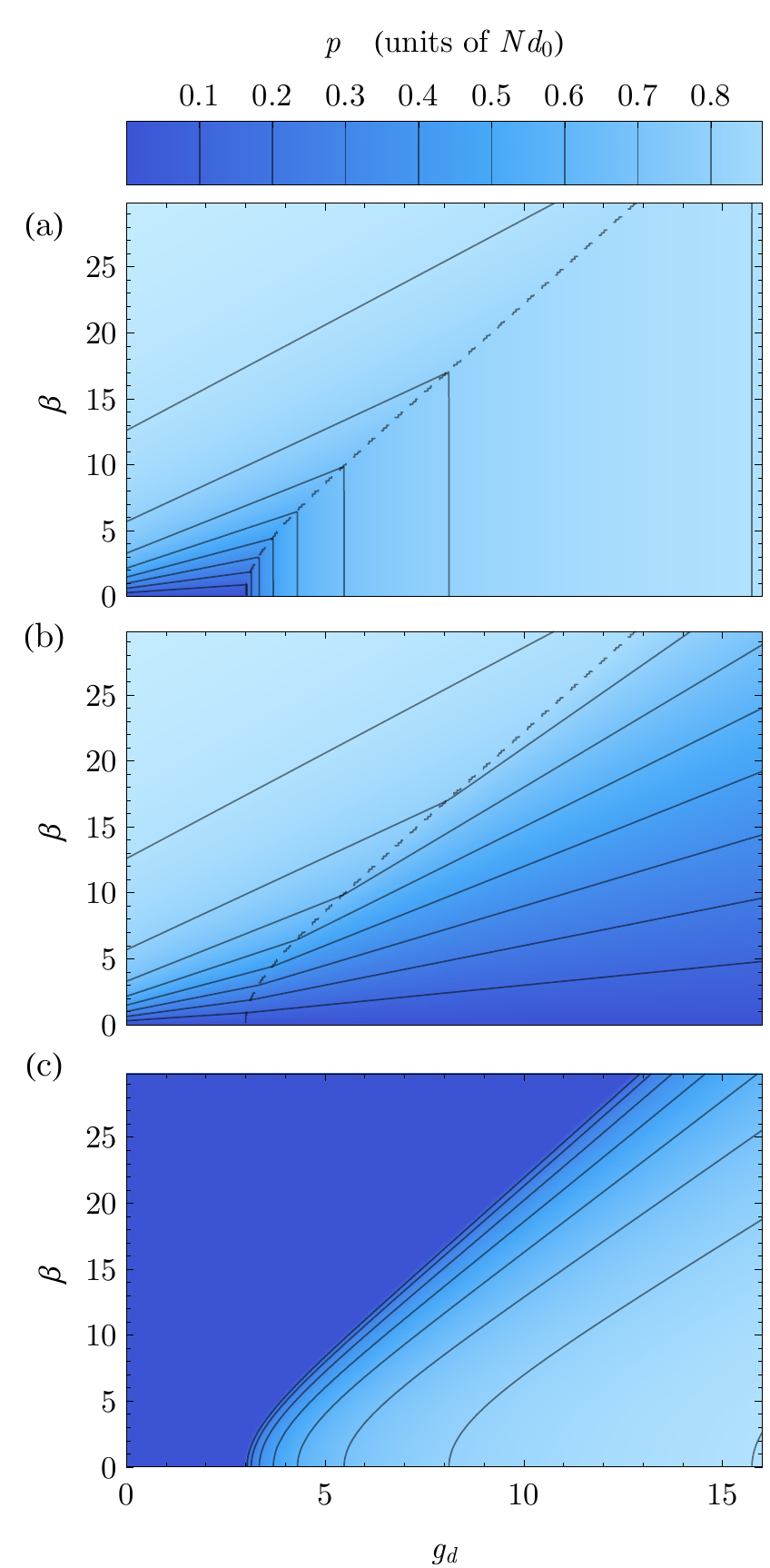}
\caption{The (a) magnitude $p$, (b) $z$-component $p_z$, and (c) in-plane component $p_{\perp}$ of the mean-field ground state polarization as a function of interaction
strength $g_{\textrm{d}}$ and electric field strength $\beta$. To the
left of the dashed lines, the in-plane component $p_{\perp}$ of the
polarization is zero. (a) In the region where $p_{\perp}\neq0$, $p$ is constant with $\beta$, indicating that with increasing field
strength, the molecules \emph{tilt} rather than polarize. (b) The $z$-component $p_z$ of the polarization increases with $\beta$;
increasing $g_\text{d}$ drives the system toward zero polarization due to
the repulsive nature of the interactions. (c) The in-plane component $%
p_{\perp}$ is uniformly zero for small-enough values of $g_\text{d}$, but a
non-zero $p_{\perp}$ develops at a field-dependent critical interaction
strength, rising continuously from zero. }
\label{fig:dip1}
\end{figure}

\begin{figure}[t!]
\includegraphics[width=86mm]{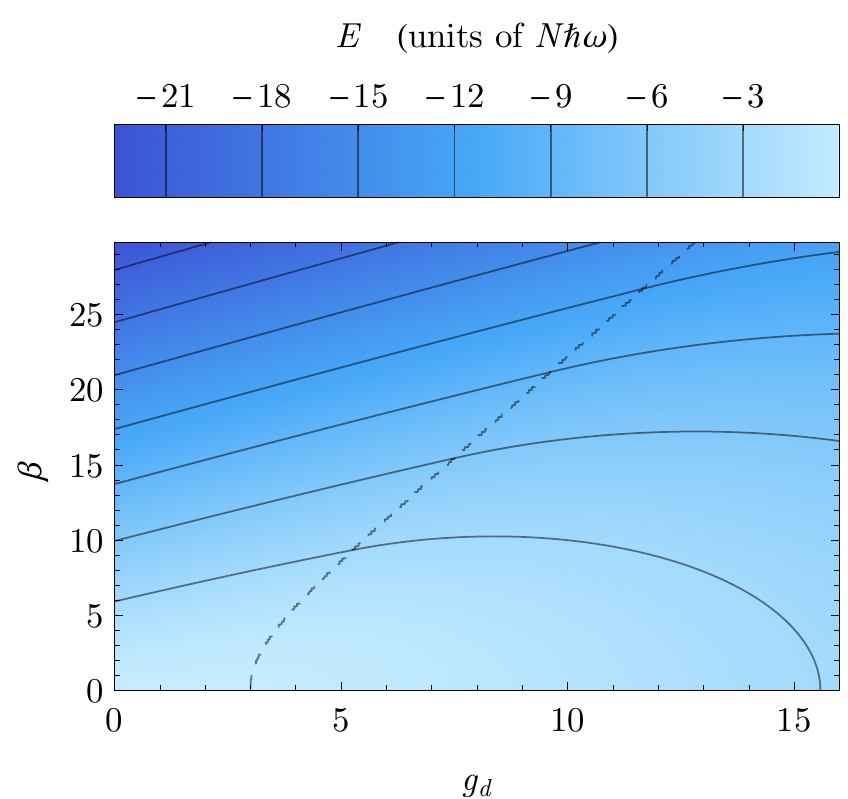}
\caption{Mean-field ground state energy per particle as a function of the
interaction strength $\beta$ and the interaction strength $g_{\text{d%
}}$. As the electric field strength increases, the energy decreases. In the
limit of zero interactions, the energy displays a quadratic Stark shift near
zero field and a linear Stark shift for larger applied fields. For fixed
field, the energy increases as a function of interaction strength until a
field-dependent critical threshold is reached, at which point the mean field
energy decreases with interaction strength.}
\label{fig:gse}
\end{figure}

In Fig.~\ref{fig:dip1}, we plot different components of the bulk
polarization of the gas as a function of $\beta$ and $g_{\text{d}}$. Plotted
in the Fig.~\ref{fig:dip1}(a) is the magnitude $p$ of the polarization $\vec{p}$. In the region
to the left of the dashed curve, $p$ increases as the electric field
increases, indicating that the gas is being polarized by the external field.
In addition, as $g_\text{d}$ increases, the polarization decreases; stronger
interactions means that the local field sampled by each molecule is smaller,
since the dipole field of the other molecules is locally anti-aligned with the
external field at the position each molecule.
In the region to the right of the dashed curve, $p$ is
constant with $\beta$, while $p$ increases with $g_\text{d}$. The threshold
value $g_{\text{c}}$ of the interaction strength that demarcates these two
regions is field-dependent and is represented by the dashed line on Figs.~%
\ref{fig:dip1}(a,b) and Fig.~\ref{fig:gse}.

The components of $\vec{p}$ provide insight into this behavior. We have
plotted in Fig.~\ref{fig:dip1}(b) the component $p_{z}$ of the polarization
parallel to the electric field and in Fig.~\ref{fig:dip1}(c) the component $%
p_{\perp}$ of the polarization perpendicular to the electric field. When $g_{%
\text{d}}<g_{\text{c}}$, $p_{\perp} = 0$ so that $p = p_{z}$. In this
region, the polarization is aligned with the external field, and the
molecules are polarized in the usual way. When $g_{\text{d}}>g_{\text{c}}$, $%
p_{\perp}$ is non-zero, and so the polarization develops a tilt into the
trapping plane. As noted before, in this region, $p$ remains constant as $%
\beta$ increases, but the magnitude of the tilt (i.e.~$p_{\perp}$)
decreases, indicating that increasing the field strength tends to align the
dipole moments with the external field. Finally, $p_{\perp}$ grows with $g_%
\text{d}$ while $p_z$ decreases, indicating that increasing the interaction
strength tilts $\vec{p}$ into the plane without changing its magnitude.

\begin{figure}[t!]
\includegraphics[width=86mm]{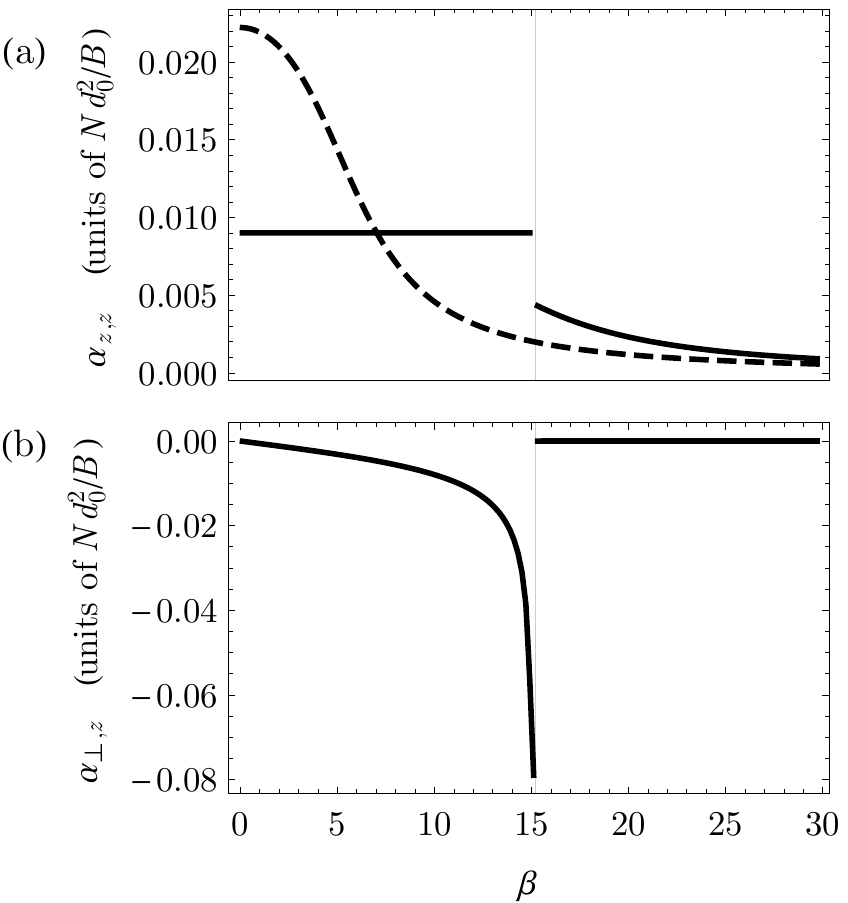}
\caption{Components of the polarizability tensor of the mean field ground
state when $p_{\perp}=0$ at $g_\text{d}=3$ (dashed) and when $p_{\perp}\neq0$ at $g_\text{d}=7.5$ (solid). (a) In the
region to the left of the tilt threshold (dashed), the $zz$ component $\alpha%
_{zz}$ of the polarizability tensor decays smoothly to zero as the field
increases, indicating a saturation effect in which the dipoles become
maximally aligned with the field. In the region where the molecules tilt (solid), $%
\alpha_{zz}$ is constant; it is discontinuous across the threshold
and decays to zero outside the tilt region. (b) The $\perp$-$z$ component $%
\alpha_{\perp z}$ of the polarizability is zero at zero field,
indicating that the strong interactions cause tilted dipoles to spontaneously
form. The polarizability decreases and diverges as the field is increased
toward the tilt threshold, since as the field increases, dipole moments
aligned with the field become more favored.}
\label{fig:polar1}
\end{figure}

To further clarify this behavior, we investigate components of the polarizability tensor $\boldsymbol{\alpha}$, given by 
\begin{align*}
\alpha_{zz} &= \frac{\partial p_z}{\partial z}, \\
\alpha_{\perp,z} &= \frac{\partial p_{\perp}}{\partial z}.
\end{align*}
We have plotted $\alpha_{zz}$ and $\alpha_{\perp,z}$ as a function of $\beta$
in Fig.~\ref{fig:polar1}. In Fig~\ref{fig:polar1}(a) the $zz$ component $%
\alpha_{zz}$ is plotted for $g_{\text{d}}=3$ (dashed) and $g_{\text{d}}=7.5$
(solid). When $g_{\text{d}}=3$, $\alpha_{zz}$ decays smoothly towards zero as the
molecules become maximally polarized in the direction of the electric field.
In the case where $g_{\text{d}}=7.5$, $a_{zz}$ is constant with increasing
electric field strength, until $\beta\approx15$ after which there is a
discontinuous jump in $\alpha_{zz}$ which then decays smoothly to zero. The
discontinuity occurs as the electric field strength is increased across the
tilt threshold, causing the dipole moments to align with the electric field.
The polarizability then decays smoothly towards zero as the molecules become
maximally polarized in the direction of the electric field.

In Fig~\ref{fig:polar1}(b) we plot $\alpha_{\perp,z}$ for $g_{\text{d}}=7.5$. When $g_{\text{d}}=7.5$, $\alpha_{\perp z}$ is zero at zero field, indicating
that the external field is too weak to overcome the attractive
interactions between tilted dipoles. The polarizability decreases as $\beta$
increases and diverges as $\beta\to15$, which occurs as the dipole
moments become aligned with the electric field. For values of $\beta$ above
this threshold, $p_{\perp} = 0$ and so $\alpha_{\perp z}=0$.

Finally, in Fig.~\ref{fig:gse} we plot the energy of the mean-field ground
state as a function of $\beta$ and $g_{\text{d}}$. In the regime where $g_{%
\text{d}}$ is significantly less than $g_{\text{c}}$, the energy decreases
linearly with $\beta$, indicating a linear Stark shift. In the region where $%
g_{\text{d}}<g_{\text{c}}$, the energy increases with $g_{\text{d}}$, since
the ddi is repulsive when the dipole moments are aligned with the external
field. In contrast, when $g_{\text{d}}$ is much larger than $g_{\text{c}}$,
the energy begins to decrease with $g_\text{d}$, meaning that the dipole
moments have tilted enough so that the ddi is attractive.

From the preceding, a picture of the behavior of the mean-field ground state
takes shape. This behavior is driven by three competing mechanisms.
First, the external field acts to align the molecular dipole moments and stretch them along the field axis ($\hat{\mathbf{z}}$). The extent to which the molecules are polarized depends on the ratio of field strength $dE$ to rotational constant $B$, since $B$ determines the zero-field splitting of the molecular rotational states. Second, as long as the dipoles are aligned with the external field, the dipole-dipole interactions act to reduce ${p}$, since the dipole field to due to all other molecules is locally anti-aligned with the external field at the position of each molecule. Finally, there are two ways in which the $p_z$-component of the dipole moment can be reduced, driven by the ddi. If $\vec{p}$ is aligned with the external field, then either $\vec{p}$ shrinks or $\vec{p}$ flips direction. Alternatively, the magnitude of the dipole moment $p$ can remain constant while the dipole tilts away from the field/trap axis.

The competition between the alignment of the dipoles along the field axis and the tilting of the dipoles away from the trap axis manifests in the following way.
When $g_{\text{d}}<g_{\text{c}}$, the gas is uniformly
polarized in the direction of the applied electric field. Increasing the
interaction strength leads to an increase in the ground-state energy since
the interactions between molecules are repulsive. Consequently, the
polarization decreases due to the interplay between local electric fields
created by the individual molecules in the gas and the applied external
electric field. In contrast, when the interaction is strong enough, i.e.~$g_{\text{d}}>g_{\text{c}}$, the molecules ``tilt'' as they develop a component of the polarization in the trapping plane. This occurs because decreasing the
interaction energy by tilting is energetically favored over decreasing the
interaction energy by reducing the dipole moment.

At the mean-field level, the gas is stable in the
ground state. However, mesoscopic fluctuations in both the density and
polarization induced by quasi-particle excitations can cause the gas to
destabilize. It is well-known that a fully-polarized dipolar gas in quasi-2D
will go unstable at a critical interaction strength due to localized density
fluctuations \cite{Ronen06,Wilson2008a,Bohn09b,Ticknor11,Bisset13} caused by local sampling of the attractive
part of the dipole-dipole interaction. In addition, in the weak
external-field limit, a BEC of polarizable molecules can go unstable due to
localized fluctuations in the polarization \cite{Wilson14,Peden2015} caused by
attractive interactions between oppositely-aligned dipole moments. In
contrast, here, attractive interactions between tilted dipole moments can
cause the BEC to go unstable, and this manifests as a new instability at low
field associated with the tilting of the dipole moments into the trapping
plane. In order to fully understand the low-energy behavior and stability of
the gas, we need to investigate the properties of the low-energy excitations
of the gas, which is the subject of the next section.

\section{Low Energy Excitations}

\label{sec:LEE}

In this section, we analyze the low-energy excitations by way of their dispersion
relations. Using the
methods outlined in Sec.~\ref{sec:theo} and App.~\ref{app:Bogoliubov}, we
obtain the quasi-particle dispersion relations $\Omega_{k}$. The gas is
unstable when the lowest branch of $\Omega_{k}$ is complex at some finite
momentum $k$, giving us a simple way to determine the stability threshold of
the gas in $\beta$-$g_{\text{d}}$ parameter space. Finally, in the next section, we characterize
the low-energy fluctuations and the associated instabilities via static
structure factors and identify a new instability associated with the tilting
of the polarization into the trapping plane.

\begin{figure}[t!]
\includegraphics[width=8.6cm]{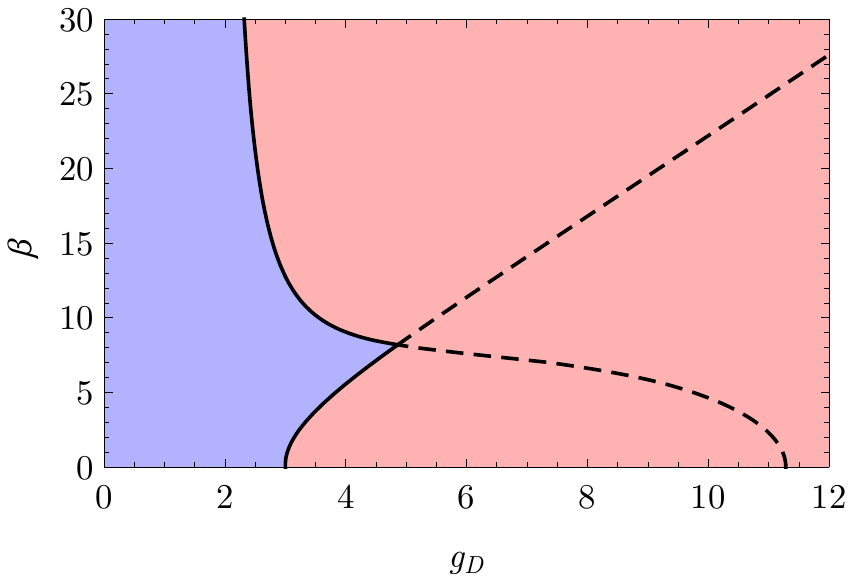}
\caption{(Color online.) Stability diagram for $B=\hbar\omega$. Red corresponds to $%
\beta$ and $g_{\text{d}}$ at which the gas will become unstable, and
blue corresponds to those that are stable. For small $\beta
$, the stability threshold is dictated by the emergence of the in-plane
polarization and tilt instability. For large $\beta$, the instability corresponds to
the softening of the density-wave roton.}
\label{fig:stabl}
\end{figure}

The stability diagram is shown in Fig.~\ref{fig:stabl}. The solid black
curve is the stability threshold, and the gas is stable for values of $\beta$
and $g_\text{d}$ in the blue region (to the left of the solid line). In the high-field limit ($\beta\gtrsim25
$), the gas goes unstable via the well-known density-wave rotonization
predicted previously. The continuation of the upper portion of
the stability threshold to low field is achieved through an artificial
restriction of the angular momentum manifold to just $m=0$. This instability
is associated with the onset of a polarization wave (see Ref.~\cite{Wilson14,Peden2015}). While not relevant for a gas of bare rigid rotor molecules, it is possible that such an approximation is relevant for microwave-dressed
molecules. This is the subject of ongoing research.

The polarization-wave instability is not present for the case of general
rigid rotor molecules. Instead, as we can see in Fig.~\ref{fig:stabl}, the
stability threshold at low field ($\beta\lesssim9$) occurs at values of $%
g_\text{d}$ much smaller than that of the polarization instability. Instead, the stability
threshold at low field and the continuation of this curve into the high
field regime is the same curve as the tilt threshold shown as a dashed curve
in Figs.~\ref{fig:dip1} and \ref{fig:gse}. This indicates that the tilting
of the polarization into the trapping plane will cause the gas to
destabilize well before the polarization instability sets in.

\begin{figure}[t!]
\includegraphics[width=8.6cm]{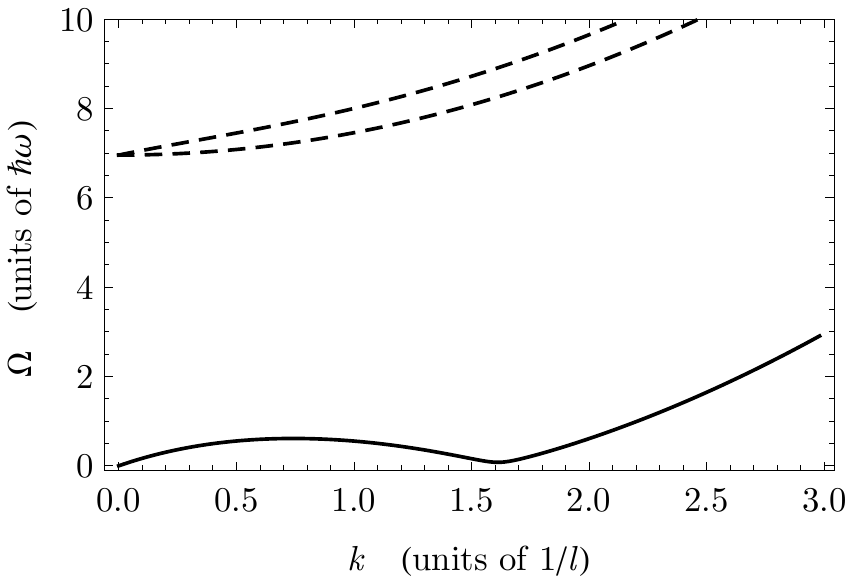}
\caption{Lowest three branches of the dispersion relation for $b=\hbar%
\omega$, $\beta = 30$, and $g_\text{d} = 2.318$, near the
stability threshold. All upper branches are indistinguishable from
free-particle dispersions. The lowest branch displays the well-known
maxon-roton feature characteristic of the density-wave instability of a
fully polarized BEC.}
\label{fig:hfe}
\end{figure}

To characterize these instabilities, we examine the dispersion relation $%
\Omega_{k}$ near the stability threshold. In the high field limit, the gas
is fully polarized in the $z$-direction and we see the emergence of a
roton-like feature in the lowest branch. This can be seen in Fig.~\ref%
{fig:hfe}, in which we plot the lowest three branches of $\Omega_k$ for $%
\beta=30$ and $g_{\text{d}}=2.81$. The lowest branch has a roton feature at $%
kl\approx1.6$. Additionally there is a large separation in energy between
the lowest branch and the upper three branches, indicating that the lowest
branch governs the behavior of the low-energy fluctuations above the
mean-field ground state. As the value of $g_{\text{d}}$ increases, the roton
feature softens and leads to the well-known density-wave instability~\cite%
{Wilson2008a}.

\begin{figure}[t!]
\includegraphics[width=8.6cm]{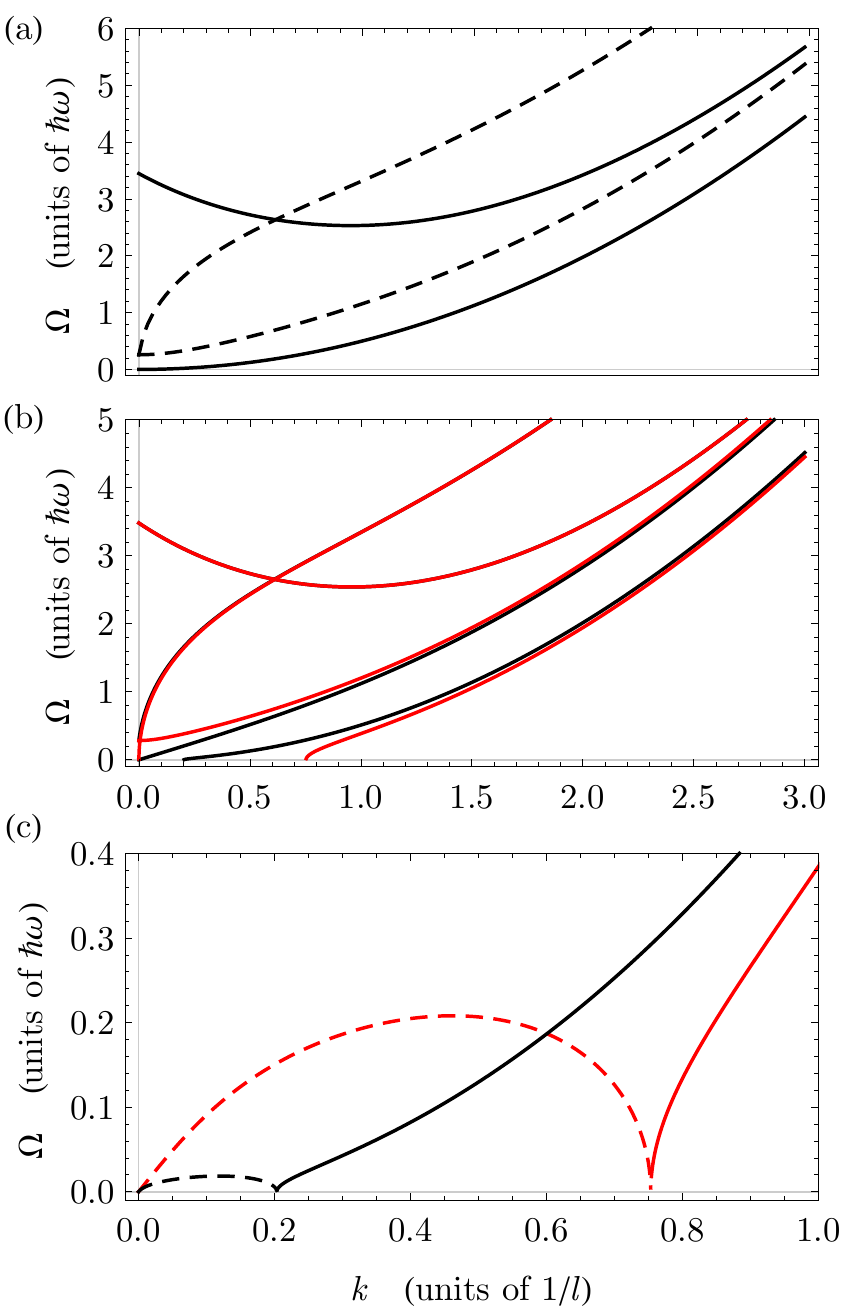}
\caption{(Color online.) Low-field dispersion relations for $\beta = 0
$, (a) $g_\text{d} = 2.95$ and (b,c) $g_\text{d} = 3.03$. (a) In the stable
regime near the instability threshold, the lowest branch (solid) is a
free-particle dispersion, and the upper branch (solid) at low momentum has
spin-wave character. The other two low-energy branches (dashed) are due to
the $m\neq0$ states. All higher-energy branches of the dispersion are
indistinguishable from free-particle dispersion relations. (b) In the
unstable regime near the threshold, the low-energy branches become
anisotropic due to the development of an in-plane component of the
polarization. The system goes unstable perpendicular to the tilt (red),
earlier than than parallel to the tilt (black). (c) The instabilities,
signaled by an imaginary component of the low-energy dispersion relations
(dashed), are infinite wavelength in nature, indicating that the instability
is global.}
\label{fig:lfe}
\end{figure}

In contrast to the high field limit, in the low field limit, the gas has a
uniformly small $z$-polarization and the roton feature is no longer present.
In Fig.~\ref{fig:lfe}, we plot the lowest four branches of $\Omega_k$ for $%
\beta=0$, and (a) $g_{\text{d}}=2.95$, (b,c) $g_{\text{d}}=3.03$. When the
interaction strength of the gas is small (Fig.~\ref{fig:lfe}(a)), the lowest
branch (solid) is a free particle dispersion and the upper branch (solid)
exhibits spin-wave character at low-momentum. The two middle branches
(dashed) correspond to $m\neq0$ molecular states.

As $g_{\text{d}}$ increases, the $m\neq0$ branches decrease in energy,
eventually leading to complex components in the dispersion relation, as
shown in Figs.~\ref{fig:lfe}(b,c). This occurs when $g_\text{d}$ is equal to 
$g_\text{c}$, indicating that the instability is due to the molecules
tilting into the plane. As the tilt develops, the azimuthal symmetry of the
system is broken, leading to an anisotopric dispersion relation. In Fig.~\ref%
{fig:lfe}(b), the dispersion is plotted in a direction perpendicular (red) and parallel
(black) to the tilt. The gas first destabilizes in the direction
perpendicular to the tilt, which can be seen more clearly in Fig.~\ref{fig:lfe}(c), where we have zoomed in on the low-$k$/low-$\Omega$ part of
the dispersion relation. The dashed curves represent the complex components
of the dispersion relation, and since they appear at $k=0$ first, the instability has an
infinite-wavelength character, indicating the presence of a global phonon-like
instability associated with the tilting of the polarization into the
trapping plane.

\begin{figure}[t!]
\includegraphics[width=8.6cm]{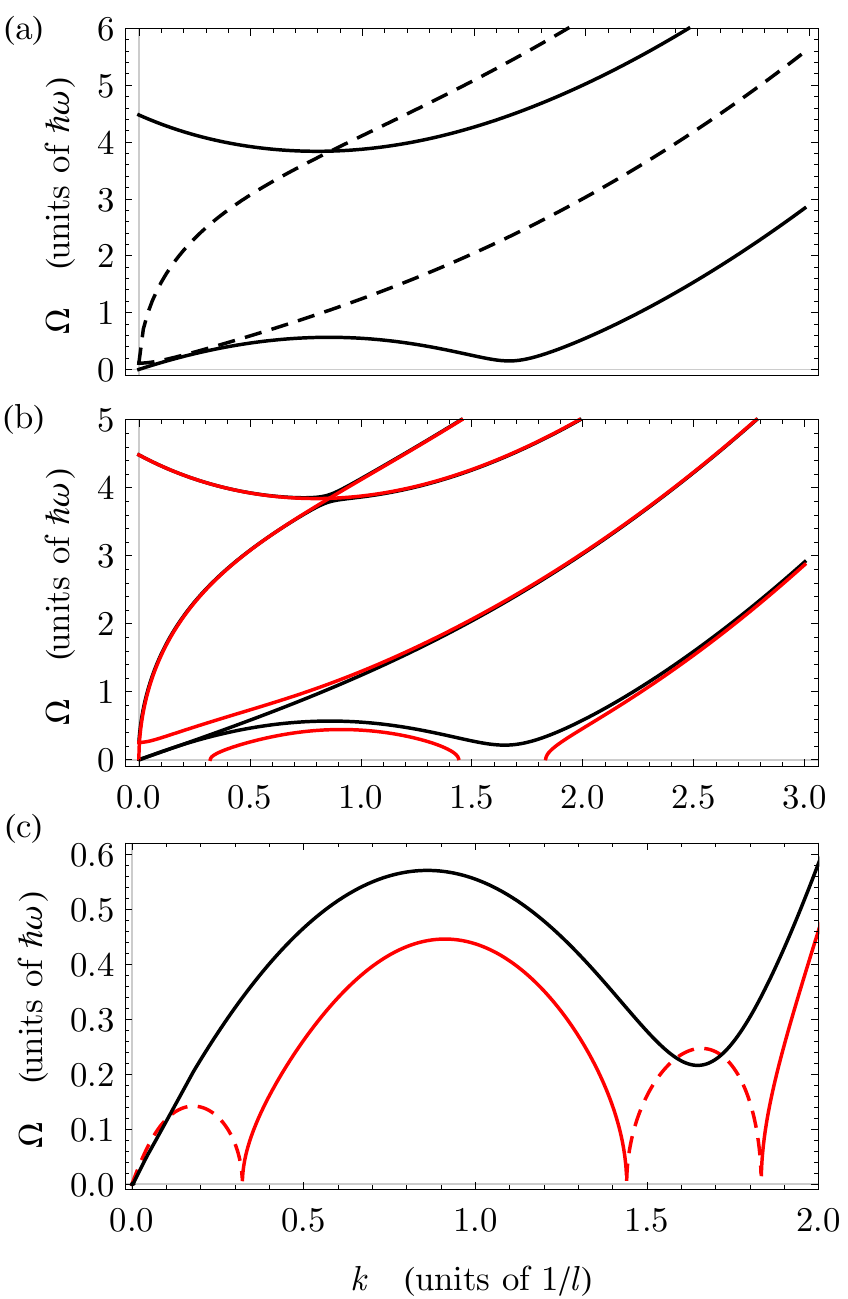}
\caption{(Color online.) Dispersion relations near the cusp between the two
instability thresholds for $\beta = 8.1$, (a) $g_\text{d} = 4.85$,
and (b,c) $g_\text{d} = 4.87$. (a) In the stable regime near the instability
threshold, the lowest branch (solid) displays a density-wave rotonization.
The two branches associated with the global tilt instability dip down to
zero energy at $kl=0$. (b) In the unstable regime near the threshold, the
low-energy branches are anisotropic. The system is unstable perpendicular to
the tilt (red) but not parallel to the tilt (black). (c) The instability has
both global and local character due to the global tilt instability at $kl=0$
and the mesoscopic density wave instability near $kl=1.5$.}
\label{fig:mfe}
\end{figure}

Finally, there is a point in parameter space where the stability of the gas
transitions from being governed by the softening of the density-wave roton
feature, seen in the high field limit, to being governed by the development
of the infinite-wavelength tilt instability, seen in the low field limit.
These two thresholds meet at a cusp. In Fig.~\ref{fig:mfe}(a), we have
plotted the lowest four branches of $\Omega_k$ for $\beta=8.1$ and $g_{\text{%
d}}=4.85$ near this cusp. In the lowest branch there is the development of a
roton-maxon feature at $kl\approx1.6$, while the zero-momentum gap of the
two branches corresponding to $m\neq0$ molecular states (dashed curves) is
very small. As $g_{\text{d}}$ increases, the dispersion relation again
becomes anisotropic (Fig.~\ref{fig:mfe}(b)), indicating that the azimuthal
symmetry of the system has been broken due to the tilting of the
polarization into the trapping plane. In addition, as seen in Fig.~\ref%
{fig:mfe}(c), there are complex components (dashed curves) in the dispersion
relation perpendicular to the tilt (red), while there are no complex
components parallel the tilt (black). Finally, we can see that both
phonon-like and density-wave instabilities are present, since the dispersion
relation has gone complex both at zero-momentum and at the roton feature
near $kl\approx1.6$.

In our analysis, we see two ways in which the BEC ground state destabilizes.
In the limit of a strongly polarizing electric field, the instability
develops due to the softening of a roton-maxon feature in the lowest branch
of the dispersion. For small applied field, the instability develops due to
the dispersion relation going complex at zero-momentum, with the dispersion
becoming anisotropic due to the tilting of the polarization into the
trapping plane. In the next section we will investigate the physical
character of the gas near the stability threshold in both the high- and low-
field limits, allowing us to characterize the mechanisms giving rise to
these instabilities.

\section{Physical characterization of the instabilities}
\label{sec:physicalcharacterization}

In the previous section, we identified two features in the dispersion
relations related to distinct ways in which the gas goes unstable. In the
high-field limit, softening of a roton at finite momentum in the lowest
branch of the dispersion causes the gas to go unstable. In contrast, in the
low-field limit, the branches corresponding to the $m\neq0$ molecular states
decrease in energy as the interaction strength increases and become complex
at zero momentum. The former instability corresponds to the well-known
density-wave instability that occurs in a fully polarized BEC, but the
latter is a new phenomenon associated with the spontaneous symmetry breaking
that occurs when the polarization tilts into the trapping plane.

In order to fully characterize the physical nature of the instabilities, we
examine the momentum-space density-density and spin-spin structure factors
(Eqs.~\ref{eqn:StructureFactor} and \ref{eqn:StructureFactorPerp}), which encode information about the spontaneous fluctuations that arise above the mean-field ground state.
In particular, we will examine the density-density structure factor $s_n$, the
spin-spin structure factor $s_z$ corresponding to the $z$-component of the
polarization, and the spin-spin structure factor $s_{\perp}$ corresponding
to the in-plane components of the polarization.

\begin{figure}[t!]
\includegraphics[width=8.6cm]{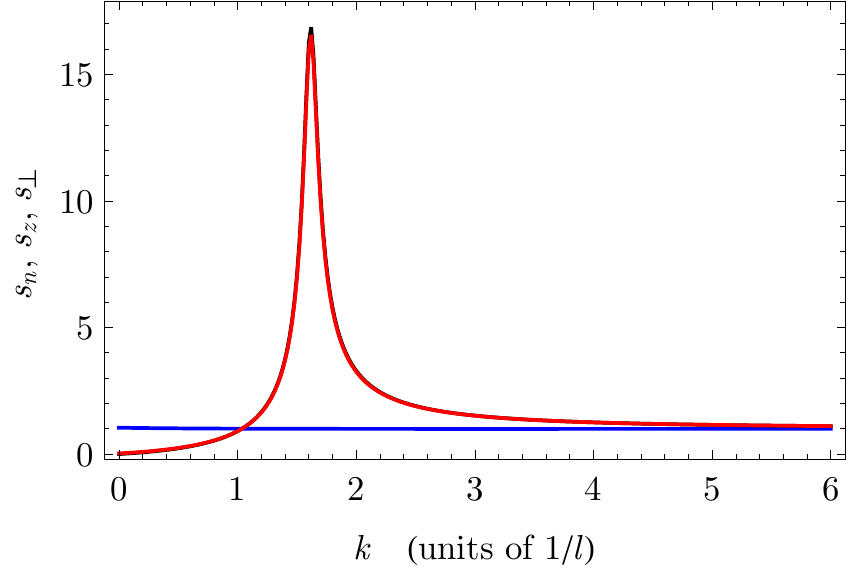} 
\caption{(Color online.) Structure factors for $\beta=30$ and $g_\text{d}=2.31$. The density structure factor $s_n$ (black) is peaked at a
momentum corresponding to the roton feature in the dispersion relation (see
Fig.~\ref{fig:hfe}) as is the $z$-component of the polarization
structure factor $s_z$ (red). The structure factor for the in-plane
component of the polarization (blue) is featureless, since this point in
parameter space is far away from the tilt instability.}
\label{fig:sfhf}
\end{figure}

In Fig.~\ref{fig:sfhf}, we plot the structure factors for $\beta=30$ and $g_{\text{d}}=2.31$, which is near the threshold of the density-wave instability. In this limit high-field limit, the gas is uniformly
polarized along the $z$ direction. Both $s_n$ and $s_z$ are strongly peaked
at $kl\approx1.6$, which is the position of the roton minimum in the dispersion relation.
These two structure factors are identical because the polarization
fluctuations (described by $s_z$) arise as a consequence of the density
fluctuations (described by $s_n$) \cite{Peden2015}. This indicates that the
gas goes unstable due to mesoscopic density fluctuations. There are no
features in $s_{\perp}$, indicating that there are no fluctuations
associated with $d_{\perp}$.

\begin{figure}[t!]
\includegraphics[width=8.6cm]{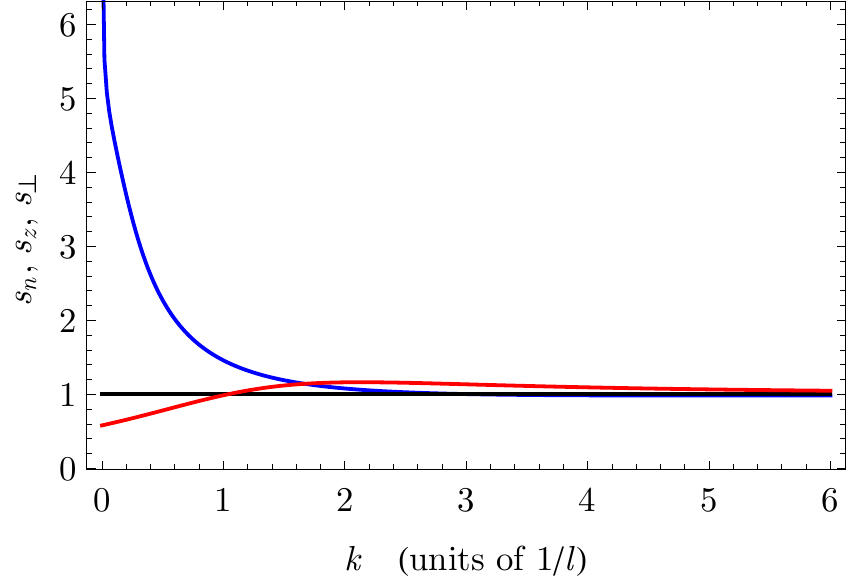}
\caption{(Color online.) Structure factors for $\beta=0$ and $g_{%
\text{d}}=2.95$. The density structure factor $s_{n}$ (black) is flat due to
the absence of density fluctuations in the low-field limit. The structure
factor $s_{z}$ (red) corresponding to the $z$-component of the polarization
displays a small peak due to spin-wave fluctuations. The structure factor
for the in-plane component $s_{\perp}$ (blue) of the polarization, develops
a peak at $k=0$ that diverges as $g_\text{d}\to3$, indicating the existence
of a global phonon-like instability.}
\label{fig:sflf}
\end{figure}

In Fig.~\ref{fig:sflf}, we plot the structure factors for $\beta=0$ and $g_{%
\text{d}}=2.95$, a point near the low-field threshold corresponding to the tilt instability. In this low field limit, the mean-field polarization is nearly zero. However,
for sufficiently large values of $g_{\text{d}}$, a non-zero in-plane
component of the polarization develops as the molecules tilt into the
trapping plane. The tilt breaks the azimuthal symmetry of the system, and
causes the gas to destabilize anisotropically near $kl=0$ (see Fig.~\ref%
{fig:lfe}). The structure factor $s_{\perp}$ diverges at $kl=0$, indicating
that the instability in this regime is infinite-wavelength (global) in
nature and therefore phonon-like. The density-density structure factor $s_{n}
$ is featureless, indicating that the gas is not susceptible to density-wave
fluctuations. This is further confirmed by the absence of a roton feature in
the lowest branch of the dispersion (see Fig.~\ref{fig:lfe}). The gas is
weakly susceptible to fluctuations in $d_z$, as seen in Fig.~\ref{fig:sflf},
where $s_{z}$ is not completely featureless. This is an artifact of the
spin-wave instability seen in a multi-state dipolar BEC with an angular
momentum manifold restricted to just $m=0$ (see Ref.~\cite{Peden2015}). Here,
the new tilt instability occurs for values of $g_{\text{d}}$ much smaller than that
required to see the spin-wave instability.

\begin{figure}[t!]
\includegraphics[width=8.6cm]{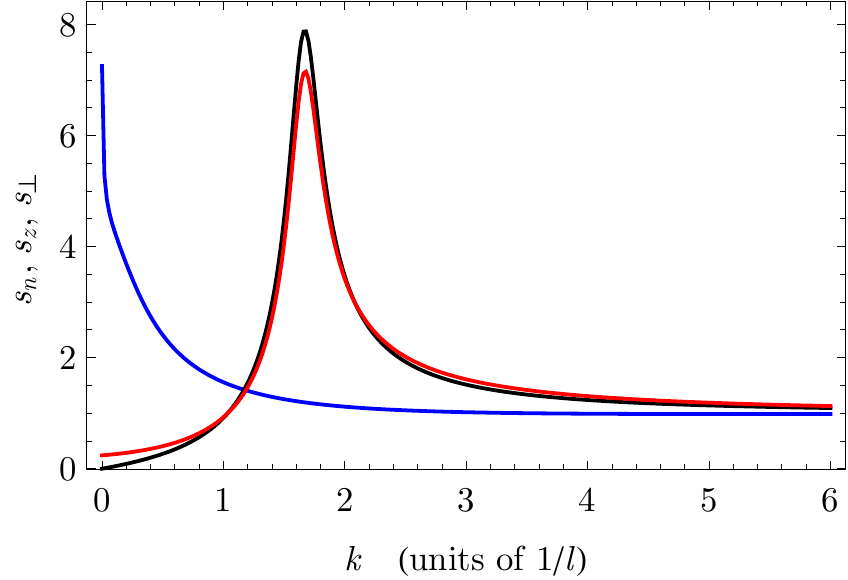}
\caption{(Color online.) Structure factors for $\beta=8.1$ and $g_{%
\text{d}} = 4.8$, near the cusp between the two different instability
thresholds. The structure factors for both the density and $z$-component of
polarization (respectively, $s_{n}$ (black) and $s_z$ (red) are peaked at
the momentum which corresponds to the roton feature in the dispersion
relation (see Fig.~\ref{fig:mfe}). The structure factor $s_{\perp}$
(blue) for the in-plane component of the polarization develops a peak at $k=0
$.}
\label{fig:sfmf}
\end{figure}

Finally, in Fig.~\ref{fig:sfmf}, we plot the structure factors for $\beta=8.1
$ and $g_{\text{d}}=4.8$ near the cusp between the density-wave and tilt
instability thresholds. Strong peaks develop in both $s_n$ and $s_{\perp}$
near the stability threshold, indicating that the gas is displaying
behaviors characteristic of both the density-wave and tilt instabilities.

From Figs.~\ref{fig:sfhf}, \ref{fig:sflf}, and \ref{fig:sfmf}, a clear
picture of the ways in which the instabilities arise in the condensate
emerges. In the high field limit, the stability of the gas is determined by a
density wave rotonization. The modulations in the density are a result of
attractive tip-to-tail interactions caused by molecules moving in the axial
direction of the trapping potential. The periodic domains of high and low
density trivially lead to periodic domains of high and low polarization,
which is why there is both a density and polarization wave present in this
limit. For a more thorough discussion of this instability, see Refs.~\cite%
{Peden2015,Bohn09b}.

In the low-field limit, the instability is a result of the attractive
tip-to-tail dipole-dipole interactions which occur as the molecules spontaneously tilt into the
trapping plane. This tilt breaks the azimuthal symmetry of the system, which
leads to an anisotropic dispersion, with the gas destabilizing in the
direction perpendicular to the tilt first. The phonon-like instability seen
in this regime replaces the spin-wave instability that arises in a dipolar condensate with
an angular momentum manifold restricted to just $m=0$ states, previously predicted in Refs.~\cite{Wilson14,Peden2015}.

\section{Conclusion}\label{sec:Conclusion}

In this paper we consider the behavior of the mean field ground state and low-energy excitations of a quasi-2D molecular BEC consisting of rigid rotor molecules in the presence of an external electric field oriented parallel to the trap axis. Due to the spin-exchange between internal and external degrees of freedom induced by the dipole-dipole interactions, the system undergoes a second-order phase transition in the mean-field ground state at low fields. This phase transition corresponds to the spontaneous symmetry-breaking of the azimuthal symmetry about the external electric field, occurring as the net dipole moment of the gas ``tilts'' and develops an in-plane component.

The system immediately goes unstable across this transition because as soon as the dipole moments develop a tilt, they can sample the attractive tip-to-tail part of the dipole-dipole interaction. This behavior is in contrast to what occurs in the high-field limit in which the system goes unstable towards the well-known density-wave rotonization that occurs in a fully polarized gas. There, the gas goes unstable once it becomes favorable energetically for the molecules to move to higher values of the trapping potential in order to sample the attractive part of the dipole-dipole interaction. This ``barrier'' caused by the trapping potential isn't relevant when the molecules tilt off the trapping axis.

Finally, due to the breaking of the azimuthal symmetry when the system tilts, the properties of the system become anisotropic. This is evidenced by the dispersion relations (Figs.~\ref{fig:lfe} and \ref{fig:mfe}) which differ along different directions as soon as the the gas enters the tilt phase. While the gas is unstable in this regime, the instability is phonon-like in the sense that it is a long-wavelength phenomenon, as evidenced by the peak in the in-plane polarization structure factor at $k=0$ (see Fig.~\ref{fig:sflf}). Such a long-wavelength instability might be stabilized in a fully trapped system in which the long-wavelength behavior is cut off due to the finite size of the gas. In addition, the last term in Eq.~\ref{eqn:integratedddionoperator} allows for exchange of angular momentum between the internal and center-of-mass degrees of freedom, indicating that this tilt may influence the center-of-mass structure of the ground state in a trapped system. As a consequence, the tilt instability coupled with the spin exchange might manifest as spin textures in the ground state in a trapped system. We have conducted some preliminary research which suggests that this occurs, but we leave this for future work.

Finally, in the regions of parameter space near the instabilities, the dipole-dipole interactions dominate the behavior of the system. In this case, beyond-mean-field physics are important. In particular, the LHY correction~\cite{Lee1957,lima2012}, which we've neglected, can stabilize a phase in which the BEC breaks up into droplets. Novel phenomena arise in such cases, such as supersolidity~\cite{Schmitt2016,Bottcher2019,Ilzhofer2019,Blakie2020}, and it is possible that the dynamic instabilities that we have investigated in this paper might manifest.

\begin{figure}[t!]
\includegraphics[width=8.5cm]{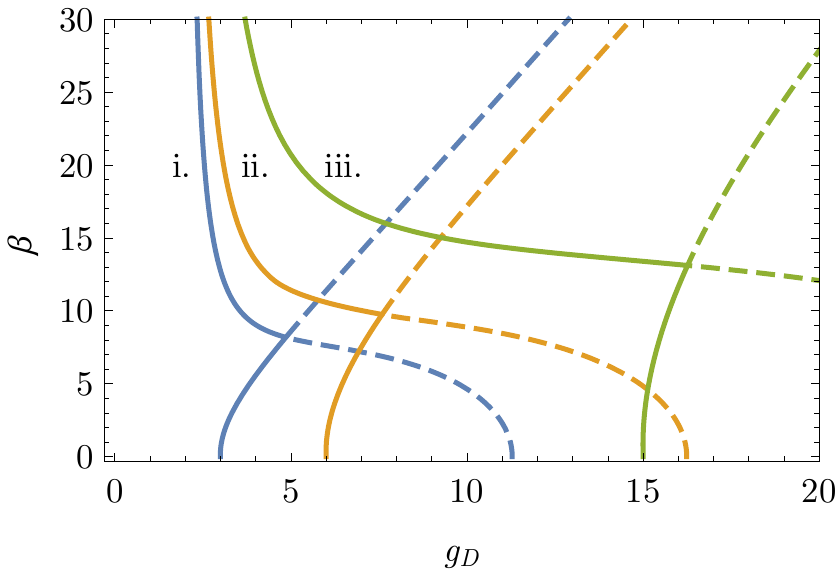} 
\caption{(Color online.) Stability thresholds for i.~$B=\hbar\omega$, ii.~$B=2\hbar\omega$, and iii.~$B = 5\hbar\omega$. In the high-field limit, all of these thresholds approach the same vertical asymptote, since the high-field instability has a purely density-wave character, and hence the splitting between internal molecular states is irrelevant.  In contrast, the splitting dominates the physics in the low field, and as a consequence the position of the stability threshold strongly depends on the rotational constant.}
\label{fig:stabilityB}
\end{figure}

\textbf{Experimental considerations.}
In the results above, we investigated a regime in which the rotational constant of the molecules is on the order of the trap frequency, i.e.~$\hbar\omega = B$. This is highly unrealistic in the context of real experiments, where the tightest traps have frequency in the 10 kHz range, whereas rotational constants for diatomic heteronuclear bialkali atoms such as RbCs are on the order of 10 GHz. Since the molecular splitting dominates the physics at low field, the position of the stability threshold is highly dependent on the rotational constant. As can be seen in Fig.~\ref{fig:stabilityB}, where we have plotted the stability diagram for three values of the ratio $B/\hbar\omega$, the threshold for the tilt-instability moves to larger values of $g_{\textrm{d}}$ as the ratio increases. The exact threshold can be determined analytically at zero-field, and it is given by
$g_{\textrm{d}} = 3{B}/{\hbar\omega}$.

For molecules such as RbCs, $B/\hbar\omega$ is on the order of $10^6$, indicating that the effective interaction strength should be on the order of $g_{\textrm{d}} \approx 3\times10^6$. Under these assumptions, the three-dimensional number density of the gas would have to be on the order of $10^{21}~\tm{cm}^{-3}$, which is orders of magnitude larger than highest densities achievable in quantum gases (for comparison, this is essentially the number density of dry air at atmospheric pressure). However, the splitting between the various levels can be tailored by applying magnetic fields, and this can reduce the value of $g_{\textrm{d}}$ required to see this instability. This is the subject of ongoing research.

\newpage \appendix

\section{Bogoliubov Theory of Rigid Rotor Molecules\label{app:Bogoliubov}}

In Sec.~\ref{sec:theo}, we outlined the process by which we arrive at the mean-field ground state energy functional, the fluctuation Hamiltonian, and the quasi-particle dispersion relations. Here, starting from Eqs.~\ref{eqn:FullManyBodyHamiltonian} and \ref{eqn:Expansion}, we fill in many of the details.  We expand the Hamiltonian to second order in the fluctuation operators $\hat{a}_{\vec{k},j}$, resulting in
\begin{align}
\hat{H}\approx K_{0}+\hat{K}_{2},
\end{align}
where%
\begin{align}
\frac{K_{0}}{N} & =-\mu\left\langle \alpha|\alpha\right\rangle +\frac {%
\hbar\omega}{2}\left\langle \alpha|\alpha\right\rangle +\left\langle
\alpha\right\vert \hat{H}_{\mathrm{mol}}\left\vert \alpha\right\rangle +%
\frac{1}{2}\left\langle \alpha\alpha\right\vert \hat{\Lambda}_{0}\left\vert
\alpha\alpha\right\rangle,  \label{eqn:GroundStateEnergyFunctionalMu} \\
\hat{K}_{2} & =\sum_{\vec{k}\neq0}\hat{\mathbf{A}}_{\vec{k}}^{\dagger}H_{%
\vec{k}}\hat{\mathbf{A}}_{\vec{k}}.  \label{eqn:FluctuationHamiltonian}
\end{align}
Here, $\hat{\Lambda}_{\vec{k}}$ is the interaction operator integrated over
spatial degrees of freedom, given by%
\begin{align}
\hat{\Lambda}_{\vec{k}}&=\frac{N}{A^{2}}\int d^{3}r_{1}\left\vert f\left(
z_{1}\right) \right\vert ^{2}e^{-i\vec{k}\cdot{\boldsymbol{\rho}}_{1}} 
\notag \\
&\quad\mbox{}\times \int d^{3}r_{2}\hat{U}\left( \vec{r}_{1}-\vec{r}%
_{2}\right) \left\vert f\left( z_{2}\right) \right\vert ^{2}e^{i\vec{k}\cdot{%
\boldsymbol{\rho}}_{2}},
\end{align}
$\mathbf{\hat{A}_{\vec{k}}^{\dagger}}$ is a row vector of creation and
annihilation operators, given by%
\begin{align}
\mathbf{\hat{A}_{\vec{k}}^{\dagger}}=\left[ 
\begin{array}{cccc}
\hat{a}_{\vec{k},1}^{\dagger} & \cdots & \hat{a}_{-\vec{k},1} & \cdots%
\end{array}
\right] ,
\end{align}
and $H_{2,\vec{k}}$ is a matrix, given by 
\begin{align}
H_{2,\vec{k}}=\left[ 
\begin{array}{cc}
\Sigma_{\vec{k}} & \Delta_{\vec{k}} \\ 
\Delta_{\vec{k}}^{\dagger} & \Sigma_{\vec{k}}^{T}%
\end{array}
\right] ,
\end{align}
where%
\begin{align}
\Sigma_{\vec{k},mn} & =\left( \frac{\hbar^{2}k^{2}}{4m}+\frac{\hbar\omega }{4%
}-\frac{\mu}{2}\right) \delta_{mn}+\frac{1}{2}\left\langle m\right\vert \hat{%
H}_{\mathrm{mol}}\left\vert n\right\rangle  \notag \\
& \quad\mbox{}+\frac{1}{2}\sum_{jk}\alpha_{j}^{\ast}\alpha_{k}\Lambda _{\vec{%
k},mjnk}+\frac{1}{2}\sum_{jk}\alpha_{j}^{\ast}\alpha_{k}\Lambda _{0,mjkn}, \\
\Delta_{\vec{k},mn} & =\frac{1}{2}\sum_{jk}\alpha_{j}\alpha_{k}\Lambda _{%
\vec{k},mnjk},
\end{align}
and%
\begin{align}
\Lambda_{\vec{k},m_{1}m_{2}n_{2}n_{1}}=\left\langle m_{1}^{(1)}\right\vert
\left\langle m_{2}^{(2)}\right\vert \hat{\Lambda}_{\vec{k}}\left\vert
n_{2}^{(2)}\right\rangle \left\vert n_{1}^{(1)}\right\rangle .
\end{align}
We note that we have ignored a term equal to $\sum_{\vec{k}}\mathrm{Tr}%
\left( \Sigma_{\vec{k}}\right) $, which is part of an LHY correction~\cite{Lee1957,lima2012} that
we ignore since we are not taking into account beyond-mean-field effects. Finally, the
chemical potential $\mu$ can be determined by minimizing Eq.~\ref%
{eqn:GroundStateEnergyFunctionalMu} with respect to the $\alpha$'s, yielding 
\begin{align}
\mu=\frac{\hbar\omega}{2}+\left\langle \alpha\right\vert \hat{H}_{\mathrm{mol%
}}\left\vert \alpha\right\rangle +\left\langle \alpha \alpha\right\vert \hat{%
\Lambda}_{0}\left\vert \alpha\alpha\right\rangle .
\end{align}

The fluctuation Hamiltonian (Eq.~\ref{eqn:FluctuationHamiltonian}) can be
diagonalized via a canonical transformation of the annihilation operators 
\cite{MingWen2009},%
\begin{align}
\hat{a}_{\vec{k},n} = \sum_{m}\left(U_{\vec{k},nm}\hat{b}_{\vec{k},n}+V_{-%
\vec{k},nm}\hat{b}_{-\vec{k},m}^{\dagger}\right),
\end{align}
where the $U$ and $V$ matrices are defined so that the matrix $X_{\vec{k}}$,
given by%
\begin{align*}
X_{\vec{k}}=\left[ 
\begin{array}{cc}
U_{\vec{k}} & -V_{-\vec{k}} \\ 
-V_{-\vec{k}}^{\ast} & U_{\vec{k}}^{\ast}%
\end{array}
\right] ,
\end{align*}
diagonalizes the matrix%
\begin{align}
H_{2,\vec{k}}\Sigma_{z}=\left[ 
\begin{array}{cc}
\Sigma_{\vec{k}} & -\Delta_{\vec{k}} \\ 
\Delta_{\vec{k}}^{\dagger} & -\Sigma_{\vec{k}}^{T}%
\end{array}
\right],
\end{align}
where
\begin{align}
\Sigma_{z}=\left[ 
\begin{array}{cc}
I & 0 \\ 
0 & -I
\end{array}
\right].
\end{align}
This results in a fluctuation Hamiltonian of the form%
\begin{align}
\hat{K}_{2}=\sum_{\vec{k},n}\frac{\Omega_{\vec{k},n}}{2}\left( \hat{b}_{\vec{%
k},n}^{\dagger}\hat{b}_{\vec{k},n}+\hat{b}_{-\vec{k},n}\hat{b}_{-\vec{k}%
,n}^{\dagger}\right) ,
\end{align}
where $\Omega_{\vec{k},n}$---the eigenvalues of $H_{2,\vec{k}}\Sigma_{z}$%
---are the quasi-particle dispersion relations.

\section{Multipole Interactions\label{app:MultipoleInteractions}}

In this appendix, we outline the derivation of $\hat{\Lambda}_{\vec{k}}$,
which is the interaction operator $\hat{U}\left( \vec{r}\right) $ averaged
over the spatial degrees of freedom, i.e.%
\begin{align}
\hat{\Lambda}_{\vec{k}}&=\frac{N}{A^{2}}\int d^{3}r_{1}\left\vert f\left(
z_{1}\right) \right\vert ^{2}e^{-i\vec{k}\cdot{\boldsymbol{\rho}}_{1}} 
\notag \\
&\quad\mbox{}\times\int d^{3}r_{2}\hat{U}\left( \vec{r}_{1}-\vec{r}%
_{2}\right) \left\vert f\left( z_{2}\right) \right\vert ^{2}e^{i\vec{k}\cdot{%
\boldsymbol{\rho}}_{2}}.
\end{align}
It can be shown that interactions between two particles that have the same
multipole $L$ can be written as 
\begin{subequations}
\begin{align}
\hat{U}_{LL}\left( \vec{r}\right) & = \mathcal{N}_{L,L}q_{L}^{2}\sum_{%
\mu=-2L}^{2L} (-1)^{\mu}\hat{T}_{2L,-\mu}\frac{Y_{2L,\mu}(\Omega_{\vec{r}})}{%
r^{2L+1}}, \\
\mathcal{N}_{L,L} &= \left(-1\right)^{L}\sqrt{\binom{4L}{2L}}\sqrt{\frac{4\pi%
}{4L+1}},
\end{align}
where $\vec{r}=\vec{r}_{1}-\vec{r}_{2}$ is the relative coordinate between
the two particles, $q_{L}$ is a matrix element in vibrational states, and $%
\hat {T}_{2L,\mu}$ is a spherical tensor operator constructed from the
single-particle multipole moments via Clebsch-Gordan coefficients.
Explicitly, these operators can be written as 
\end{subequations}
\begin{subequations}
\begin{align}
\hat{T}_{2L,\mu} & =\sum_{M=-L}^{L}\left\langle L,M;L,\mu-M|2L,\mu
\right\rangle \hat{q}_{L,M}\otimes\hat{q}_{L,\mu-M},
\label{eqn:SphericalTensorReplacements} \\
\hat{q}_{LM} & =\sqrt{\frac{4\pi}{2L+1}}Y_{LM}\left( \hat{\Omega}\right) .
\end{align}

In the following derivation, we use the following conventions. We use the
symmetric Fourier transform, given by
\end{subequations}
\begin{subequations}
\begin{align}
\mathcal{F}_{\vec{r}}\left[ f\left( \vec{r}\right) \right] \left( \vec {q}%
\right) & =\int d^{D}r\frac{e^{-i\vec{q}\cdot\vec{r}}}{\sqrt{\left(
2\pi\right) ^{D}}}f\left( \vec{r}\right) , \\
\mathcal{F}_{\vec{q}}^{-1}\left[ \phi\left( \vec{q}\right) \right] \left( 
\vec{r}\right) & =\int d^{D}q\frac{e^{i\vec{q}\cdot\vec{r}}}{\sqrt{\left(
2\pi\right) ^{D}}}\phi\left( \vec{q}\right) ,
\end{align}
in which case the convolution theorem takes the form, 
\end{subequations}
\begin{align}
\int d^{D}rf\left( \vec{r}-\vec{r}^{\prime}\right) g\left( \vec{r}\right) &=%
\sqrt{\left( 2\pi\right) ^{D}}\mathcal{F}_{\vec{q}}^{-1}[ \mathcal{F}_{\vec{r%
}}\left[ f\left( \vec{r}\right) \right] \left( \vec {q}\right)  \notag \\
&\quad\mbox{}\times\mathcal{F}_{\vec{r}}\left[ g\left( \vec{r}\right) \right]
\left( \vec{q}\right) ] \left( \vec{r}^{\prime}\right) .
\end{align}
In the case of discrete and continuous variables, respectively, the delta
function can be expanded in plane waves as 
\begin{align}
\delta_{\vec{k},\vec{k}^{\prime}} & =\int d^{2}\rho\frac{e^{i\left( \vec {k}%
^{\prime}-\vec{k}\right) \cdot{\boldsymbol{\rho}}}}{A}, \\
\delta^{D}\left( \vec{k}-\vec{q}\right) & =\int d^{D}r\frac{e^{i\left( \vec{k%
}-\vec{q}\right) \cdot\vec{r}}}{\left( 2\pi\right) ^{D}}.
\end{align}

We first write $\hat{\Lambda}_{\vec{k}}$ as%
\begin{align*}
\hat{\Lambda}_{\vec{k}}=\sum_{\mu=-2L}^{2L}(-1)^{\mu}\hat{T}%
_{2L,-\mu}\Lambda_{\vec{k},2L,\mu},
\end{align*}
where%
\begin{align}
\Lambda_{\vec{k},2L,\mu} &= \mathcal{N}_{L,L}q_{L}^{2}\frac{N}{A^{2}}\int
d^{3}r_{1}\left\vert f\left( z_{1}\right) \right\vert ^{2}e^{-i\vec{k}\cdot{%
\boldsymbol{\rho}}_{1}}  \notag \\
&\quad\mbox{}\int d^{3}r_{2}\frac{Y_{2L,\mu}(\Omega_{\vec{r}})}{r^{2L+1}}%
\left\vert f\left( z_{2}\right) \right\vert ^{2}e^{i\vec{k}\cdot{\boldsymbol{%
\rho}}_{2}}
\end{align}
\begin{widetext}
We re-write the inner integral using the convolution theorem, resulting in%
\begin{align*}
\int d^{3}x\left\vert f\left(  z\right)  \right\vert ^{2}\frac{Y_{2L,\mu
}\left(  \Omega_{\vec{r}}\right)  }{r^{2L+1}}e^{i\vec{k}\cdot{\boldsymbol{\rho}}}%
=\sqrt{\left(  2\pi\right)  ^{3}}\mathcal{F}_{\vec{q}}^{-1}\left[
\mathcal{F}_{\vec{r}}\left[  \frac{Y_{2L,\mu}\left(  \hat{r}\right)
}{r^{2L+1}}\right]  \left(  \vec{q}\right)  \mathcal{F}_{\vec{r}}\left[
\left\vert f\left(  z\right)  \right\vert ^{2}e^{i\vec{k}\cdot{\boldsymbol{\rho}}%
}\right]  \left(  \vec{q}\right)  \right]  \left(  \vec{r}\right)  .
\end{align*}
\end{widetext}
The second Fourier transform evaluates to%
\begin{align*}
\mathcal{F}_{\vec{r}}\left[ \left\vert f\left( z\right) \right\vert ^{2}e^{i%
\vec{k}\cdot{\boldsymbol{\rho}}}\right] \left( \vec{q}\right) =2\pi \mathcal{%
F}_{z}\left[ \left\vert f\left( z\right) \right\vert ^{2}\right] \left(
q_{z}\right) \delta^{2}\left( \vec{q}_{\perp}-\vec{k}\right) ,
\end{align*}
where%
\begin{align*}
\vec{q} & =\vec{q}_{\perp}+q_{z}\hat{\mathbf{z}}, \\
0 & =\hat{\mathbf{z}}\cdot\vec{q}_{\perp}.
\end{align*}
The second Fourier transform can be evaluated analytically by expanding the
exponential $e^{-i\vec{q}\cdot\vec{r}}$ in the definition of the Fourier
transform in terms of spherical harmonics. The result is 
\begin{align*}
\mathcal{F}_{\vec{r}}\left[ \frac{Y_{2L,\mu}\left( \Omega_{\vec{r}}\right) }{%
r^{2L+1}}\right] \left( \vec{q}\right) =\frac{\sqrt{2}e^{-i\pi L}}{%
2^{2L}\Gamma\left( 2L+1/2\right) }q^{2L-2}Y_{2L,\mu}\left( \Omega_{\vec{q}%
}\right) .
\end{align*}
Evaluating the integral over $\vec{q}_{\perp}$ results in 
\begin{widetext}
\begin{align*}
\int d^{3}x\left\vert f\left(  z\right)  \right\vert ^{2}e^{i\vec{k}\cdot
{\boldsymbol{\rho}}}\frac{Y_{2L,\mu}\left(  \Omega_{\vec{r}}\right)  }{r^{2L+1}}%
=\frac{2\pi\sqrt{\pi}e^{-i\pi L}}{2^{2L-1}\Gamma\left(  2L+1/2\right)
}e^{i\vec{k}\cdot{\boldsymbol{\rho}}}\int dq\frac{e^{iqz}}{\sqrt{2\pi}}\left(
k^{2}+q^{2}\right)  ^{L-1}Y_{2L,\mu}\left(  \Omega_{\vec{k}+q\hat{\mathbf{z}}}\right)
\mathcal{F}_{z}\left[  \left\vert f\left(  z\right)  \right\vert ^{2}\right]
\left(  q\right)  .
\end{align*}
Evaluating the outside integral over ${\boldsymbol{\rho}}$ in the definition of
$\Lambda_{\vec{k},L,\mu}$ yields the final, general expression,
\begin{equation}
\Lambda_{\vec{k},2L,\mu}=\frac{N}{A}\mathcal{N}_{L,L}q_{L}^{2}\frac{2\pi
\sqrt{\pi}e^{-i\pi L}}{2^{2L-1}\Gamma\left(  2L+1/2\right)  }\int dq\left(
k^{2}+q^{2}\right)  ^{L-1}Y_{2L,\mu}\left(  \Omega_{\vec{k}+q\hat{\mathbf{z}}}\right)
\mathcal{F}_{z}^{-1}\left[  \left\vert f\left(  z\right)  \right\vert
^{2}\right]  \mathcal{F}_{z}\left[  \left\vert f\left(  z\right)  \right\vert
^{2}\right]  \left(  q\right)  .
\end{equation}

When we assume Gaussian forms for the axial wave functions, i.e.%
\begin{align}
f\left(  u\right)  =\frac{1}{\sqrt{l\sqrt{\pi}}}e^{-u^{2}/2l^{2}},
\end{align}
this expression reduces to
\begin{align}
\Lambda_{\vec{k},2L,\mu}=\frac{Nq_{L}^{2}}{A}\left(  -1\right)  ^{L}%
\sqrt{\binom{4L}{2L}}\sqrt{\frac{4\pi}{4L+1}}\frac{\sqrt{\pi}e^{-i\pi L}}{2^{2L-1}%
\Gamma\left(  2L+1/2\right)  }\int dq\left(  k^{2}+q^{2}\right)
^{L-1}e^{-\left(  ql\right)  ^{2}/2}Y_{2L,\mu}\left(  \Omega_{\vec{k}+q\hat
{z}}\right)  .
\end{align}
\end{widetext}
The integrals can be evaluated analytically. To do so, we write $\vec{k}+q%
\hat{\mathbf{z}}$ as 
\begin{align*}
\vec{k}+q\hat{\mathbf{z}}=\sqrt{k^{2}+q^{2}}\left( \hat{\mathbf{x}}\sin\theta\cos\phi_{\vec{k}%
}+\hat{\mathbf{y}}\sin\theta\sin\phi_{\vec{k}}+\hat{\mathbf{z}}\cos\theta\right) ,
\end{align*}
and make the replacements%
\begin{align*}
\cos\theta & =\frac{q}{\sqrt{k^{2}+q^{2}}}, \\
\sin\theta & =\frac{k}{\sqrt{k^{2}+q^{2}}}, \\
e^{\pm i\mu\phi_{\vec{k}}} & =\left( \frac{k_{x}}{k}\pm i\frac{k_{y}}{k}%
\right) ^{\mu}.
\end{align*}

\subsection{Dipole-dipole interactions}

For dipole-dipole interactions ($L=1$), we can compute these integrals, and
the result is 
\begin{subequations}
\begin{align}
\Lambda_{2,0,\vec{k}} & =\frac{Nd^{2}}{Al}\allowbreak\allowbreak4\sqrt {%
\frac{\pi}{3}}F\left( \frac{kl}{\sqrt{2}}\right) \\
\Lambda_{2,\pm1,\vec{k}} & =0, \\
\Lambda_{2,\pm2,\vec{k}} & =\frac{Nd^{2}}{Al}\frac{2\sqrt{2\pi}}{3}%
e^{\pm2i\phi_{\vec{k}}}\left( 1-F\left( \frac{kl}{\sqrt{2}}\right) \right) ,
\end{align}
where%
\begin{equation}
F\left( x\right) =1-\frac{3}{2}\sqrt{\pi}xe^{x^{2}}\mathrm{erfc}\left(
x\right) .
\end{equation}
Upon replacing the tensor operators $\hat{T}_{2L,-\mu}$ in \ref%
{eqn:SphericalTensorReplacements}, i.e.
\end{subequations}
\begin{align*}
\hat{T}_{20} & =\frac{1}{\sqrt{6}}\left( \hat{q}_{1,1}\otimes\hat{q}_{1,-1}+2%
\hat{q}_{1,0}\otimes\hat{q}_{1,0}+\hat{q}_{1,-1}\otimes\hat{q}_{1,1}\right) ,
\\
\hat{T}_{2,\pm2} & =\hat{q}_{1,\pm1}\otimes\hat{q}_{1,\pm1},
\end{align*}
$\hat{\Lambda}_{\vec{k}}$ becomes%
\begin{align}
\hat{\Lambda}_{\vec{k}} & =g_{\text{d}}\left( \hat{q}_{1,1}\otimes\hat{q}%
_{1,-1}+2\hat{q}_{1,0}\otimes\hat{q}_{1,0}+\hat{q}_{1,-1}\otimes\hat{q}%
_{1,1}\right) F\left( \frac{kl}{\sqrt{2}}\right)  \notag \\
& +g_{\text{d}}\left( e^{-2i\phi_{\vec{k}}}\hat{q}_{1,1}\otimes\hat{q}%
_{1,1}+e^{2i\phi_{\vec{k}}}\hat{q}_{1,-1}\otimes\hat{q}_{1,-1}\right) \left(
1-F\left( \frac{kl}{\sqrt{2}}\right) \right) ,
\end{align}
where%
\begin{equation}
g_{\text{d}}=\frac{Nd^{2}\sqrt{8\pi}}{3Al}.
\end{equation}

\bibliographystyle{apsrev}
\bibliography{DipoleBEC}

\end{document}